\documentclass{PoS}
\usepackage{amsmath}
\usepackage{amsfonts}
\usepackage{multirow} 
\title{
Derivation of L\"uscher's finite size formula for $N\pi$ and $NN$ system
}
\ShortTitle{
Derivation of L\"uscher's finite size formula for $N\pi$ and $NN$ system
}
\author{
\speaker{Naruhito Ishizuka} \\
Center for Computational Science, University of Tsukuba, \\
Tsukuba, Ibaraki 305-8577, Japan \\
E-mail: \email{ishizuka@ccs.tsukuba.ac.jp}
}
\abstract{
I present derivation of L\"uscher's finite size formula
for the elastic $N\pi$ and the $NN$ scattering system 
for several angular momenta
from the relativistic quantum field theory.
}
\FullConference{
The XXVII International Symposium on Lattice Field Theory \\
26-31, 2009 \\
Peking University, Beijing, China
}
\begin{document}
%
%
\section{ Introduction }
Calculation of the scattering phase shift
represents an important step for expanding our understanding
of the strong interaction based on lattice QCD
to dynamical aspects of hadrons.
Since L\"uscher derived a finite size formula
for the two-meson system on 1986~\cite{fm_LU},
which give us a relation between the phase shift
and the energy eigenvalue on the finite volume,
many lattice calculations of the scatting length and the phase shift
of the two-meson systems have been carried with his formula.
Recently his formula was extended to that for the $N\pi$ system
by Bernard {\it et al.}
by using the non-relativistic effective theory~\cite{fm_BLMR}.
QCDSF collaboration calculated the phase shift of this system
with this extended formula
and study $\Delta(1232)$ resonance~\cite{QCDSF_Delta}.

The extension of formula
is necessary to extend our study to many systems.
In the present work I consider a derivation of the formula for 
the elastic $NN$ scattering system,
where the formula only for spin singlet state 
in the non-relativistic limit,
which is same as that for the two-meson system
given by L\"uscher,
has been known.
My derivation is based only on 
the relativistic quantum field theory
and any effective theories for the two-nucleon interaction 
are not assumed.
Further the extension to the $N\pi$ system can be easily done 
as discussed latter.
%
%
\section{ Wave function in infinite volume }
First we consider
the wave function in the infinite volume defined by 
\begin{eqnarray}
\phi^{\infty}_{\alpha\beta}({\bf x};{\bf k}) 
=
\langle 0 |
\ n_{\alpha}({\bf x}/2) 
\ p_{\beta  }(-{\bf x}/2) 
\ | {\bf k} ,   \lambda_n , \lambda_p \rangle 
\ , 
\label{eq:WF_infV_def}
\end{eqnarray}
where
$n_{\alpha}({\bf x})$ and $p_{\beta}({\bf x})$ 
are interpolating operators of the nucleons 
and $|{\bf k},\lambda_n,\lambda_p\rangle$ is 
the asymptotic $NN$ state with 
momentum ${\bf k}$, $-{\bf k}$
and the helicities $\lambda_n$, $\lambda_p$ .
Using LSZ reduction formula, 
the wave function can be written by 
\begin{eqnarray}
\phi^{\infty}_{\alpha\beta}({\bf x};{\bf k})
=
U_{\alpha\beta}({\bf k},{\lambda}_n,{\lambda}_p) 
{\rm e}^{i{\bf x}\cdot{\bf k} }
+
\int 
\frac{{\rm d}^3 p}{(2\pi)^3} 
\sum_{\xi_n \xi_p }
U_{\alpha\beta}({\bf p},\xi_n,\xi_p) 
{\rm e}^{i{\bf x}\cdot{\bf p} }
\frac{ T({\bf p},\xi_n,\xi_p;{\bf k},\lambda_n,\lambda_p) }
{ p^2 - k^2 - i\epsilon } 
\ ,
\label{eq:WF_infV_1}
\end{eqnarray}
where
$U_{\alpha\beta}({\bf k},\lambda_n,\lambda_p)$
is a spinor for two free nucleons
given by 
$U_{\alpha\beta}({\bf k},\lambda_n,\lambda_p)=
u_{\alpha}( {\bf k},\lambda_n)
u_{\beta }(-{\bf k},\lambda_p)$
with the one nucleon spinor $u({\bf k},\lambda)$.
$T({\bf p},\xi_n,\xi_p;{\bf k},\lambda_n,\lambda_p)$
is the off-shell scattering amplitude for a process
$
n( {\bf k},\lambda_n)
p(-{\bf k},\lambda_p) \to \
n( {\bf p},    \xi_n)
p(-{\bf p},    \xi_p)
$.

We can estimate
(\ref{eq:WF_infV_1})
in the region $|{\bf x}| > R$ 
for the two-nucleon interaction range $R$,
by using a integral formula
\begin{equation}
\int \frac{{\rm d}^3 p}{(2\pi)^3} 
\frac{ j_l(px) }{ p^2 - k^2 - i\epsilon } 
f(p) 
= 
\frac{k}{4\pi} 
\left( i \cdot j_l(kx) + n_l(kx) \right) f(k)      
\qquad \mbox{ for $F(x)$=0 }
\label{eq:int_formula_NI}
\end{equation}
where 
$F(x)$ is the inverse Fourier transformation of $f(p)$.
$j_l(x)$ is the spherical Bessel and
$n_l(x)$ is the Neumann function,
whose conventions agree
with those in~\cite{MESSIAH:book} as adopted in~\cite{fm_LU}.
This formula is a extension of (A.11) 
in Appendix A in Ref.~\cite{CP-PACS_wf}
to that for arbitrary value of $l$
and can be derived by similar calculations of that paper.

From (\ref{eq:int_formula_NI})
we know that
all values in the numerator of the integrand in (\ref{eq:WF_infV_1})
can be replaced by the value at on-shell $p=k$.
The off-shell scattering amplitude
$T({\bf p},\xi_n,\xi_p;{\bf k},\lambda_n,\lambda_p)$
is replaced by the on-shell amplitude,  
which can be expanded as~\cite{Hamp_JW}
\begin{equation}
T(k{\bf e}_p,\xi_n,\xi_p;{\bf k},\lambda_n,\lambda_p)
=
16\pi^2 \frac{\sqrt{s}}{k}
\sum_{JM} 
{T}_{\xi_n\xi_p , \lambda_n \lambda_p}^{(J)}(k)
\cdot
N_J{}^2
\ D_{M\xi    }^{(J)}{}^{*}({\Omega}_p)
\ D_{M\lambda}^{(J)}      ({\Omega}_k)
\ , 
\label{eq:Mphys}
\end{equation}
where
$N_J=\sqrt{(2J+1)/(4\pi)}$, 
$\lambda = \lambda_n - \lambda_p$,
$    \xi =     \xi_n -     \xi_p$ and
$\sqrt{s}=2\sqrt{m^2+k^2}$.
In (\ref{eq:Mphys}) 
the helicity amplitude in the subspace of
the total energy $\sqrt{s}$
and the total angular momentum $J$
is defined by
${T}_{\xi_n\xi_p,\lambda_n\lambda_p}^{(J)}(k)
=\langle \xi_n \xi_p | \hat{T}^{(J)}(k) | \lambda_n\lambda_p\rangle$.
The function 
$D_{MM'}^{(J)} ({\Omega}_p)
=\langle JM | 
\exp(-i\alpha J_z)
\exp(-i\beta  J_y) $ $  
\exp(-i\gamma J_z) 
|JM'\rangle$
is the Wigner's D-function with the Euler angle 
$(\alpha,\beta,\gamma)=(\phi_p,\theta_p,-\phi_p)$
for momentum
${\bf p}=
( p \sin\theta_p \cos\phi_p
, p \sin\theta_p \cos\phi_p
, p \cos\theta_p )$.

Using (\ref{eq:int_formula_NI})
and (\ref{eq:Mphys}), 
we know that 
the wave function (\ref{eq:WF_infV_1}) in the region $|{\bf x}| > R$
is written by
\begin{eqnarray}
&&
\phi^{\infty}({\bf x};{\bf k}) 
=
\sum_{JM}
N_J D_{M\lambda}({\Omega}_k) \cdot
\phi^{\infty}_{JM\lambda_n\lambda_p} ({\bf x};k)
\qquad ( \lambda = \lambda_n - \lambda_p ) 
\ , 
\label{eq:WF_infV_h1}
\\
&&
\phi^{\infty}_{JM\lambda_n\lambda_p}({\bf x};k)
= 
\sum_{\xi_n \xi_p } 
\left[
{J}_{JM\xi_n\xi_p}({\bf x};k) \cdot \alpha_{\xi_n\xi_p,\lambda_n\lambda_p}^{(J)}(k) +
{N}_{JM\xi_n\xi_p}({\bf x};k) \cdot  \beta_{\xi_n\xi_p,\lambda_n\lambda_p}^{(J)}(k) 
\right]
\ . 
\label{eq:WF_infV_h2}
\end{eqnarray}
where 
$\alpha_{\xi_n\xi_p,\lambda_n\lambda_p}^{(J)}(k)
=\langle \xi_n \xi_p | 
\hat{I} + i \hat{T}^{(J)}/2 
| \lambda_n\lambda_p\rangle$
and
$\beta_{\xi_n\xi_p,\lambda_n\lambda_p}^{(J)}(k)
=\langle \xi_n \xi_p | \hat{T}^{(J)} /2 | \lambda_n\lambda_p\rangle$,
which correspond to 
$\alpha^{(l)} = \cos\delta_l \cdot {\rm exp}(i\delta_l)$ and 
$ \beta^{(l)} = \sin\delta_l \cdot {\rm exp}(i\delta_l)$  
for the two-meson system
with the scattering phase shift $\delta_l$.
In (\ref{eq:WF_infV_h2})
the function 
${J}_{JM\lambda_n\lambda_p}({\bf x};k)$
is the wave function of two free nucleons
with the total energy $\sqrt{s}$, 
the total angular momentum $JM$ and the helicity $\lambda_n\lambda_p$.
Its explicit form is given by
\begin{equation}
{J}_{JM\lambda_n\lambda_p}({\bf x};k)
=
\hat{L}(\nabla) 
\ {J}_{JM\lambda_n\lambda_p}^{\rm NR}({\bf x};k)
\ \hat{R}(\stackrel{ \leftarrow}{\nabla})
\equiv
{J}_{JM\lambda_n\lambda_p}^{\rm NR}({\bf x};k) \bigl|_{\rm R-EX}
\ \ , 
\label{eq:J_JMhh_def}
\end{equation}
where differential operators  
$\hat{L}(\nabla)$ and
$\hat{R}(\nabla)$ are defined by 
\begin{equation}
\hat{L}(\nabla)
=
\left(
\begin{array}{c}
\displaystyle 
I   \\
\displaystyle
\frac{(\sigma\cdot\nabla/i)}{E+m}  \\
\end{array}
\right) 
\ , \quad 
\hat{R}(\nabla)
=
\left(
I   \ , \
\frac{ - (\sigma^T \cdot \nabla/i)}{E+m}
\right)
\ ,
\end{equation}
with $E=\sqrt{k^2+m^2}$.
In (\ref{eq:J_JMhh_def})
the function ${J}_{JM\lambda_n\lambda_p}^{\rm NR}({\bf x};k)$ 
is $2\times 2$ non-relativistic spinor defined by 
\begin{eqnarray}
&&
{J}_{JM\lambda_n\lambda_p}^{\rm NR}({\bf x};k)
= \sum_{ls} 
{J}_{JMls}^{\rm NR}({\bf x};k) 
\cdot \langle JMls | JM\lambda_n\lambda_p \rangle
\ , 
\label{eq:J_JMls_NR_def_1}
\\
&&
{J}_{JMls}^{\rm NR}({\bf x};k)
= j_l(kx) Y_{JM}^{ls}(\Omega_x) / b_{l}(k)
\ , 
\quad 
Y_{JM}^{ls}(\Omega_x)
= \sum_{m \mu} 
Y_{lm}(\Omega_x)
\phi(s,\mu) \cdot C( lm ; s \mu ; JM )
\ , \ \ \ 
\label{eq:J_JMls_NR_def_2}
\end{eqnarray}
where
the coefficient 
$\langle JMls | JM\lambda_n\lambda_p \rangle$
is the transformation coefficient
from the helicity base to
the orbit-spin base ($(JMls)$-base)
with the angular momentum $l$ and the spin $s$~\cite{Hamp_JW},
and $C(lm;s\mu;JM)$ is the Clebsch-Gordan coefficient
for angular momentum state $|lm\rangle\otimes |s \mu\rangle$ and $|JM\rangle$.
$\Omega_x$ is the spherical coordinate for ${\bf x}$.
$b_l(k)$ is the normalization constant of the state,
which takes 
$1/b_l(k) = (4\pi)i^l \cdot (k^2+m^2)$
for the usual relativistic normalization ($u^\dagger u = 2 E$).
$\phi(s,\mu)$ is $2\times 2$ spin wave function for two spin $1/2$ particles 
with total spin $s\mu$.
The function
${N}_{JM\lambda_n\lambda_p}({\bf x};k)$ in (\ref{eq:WF_infV_h2}) 
is given by replacing $j_l(kx)$ by $n_l(kx)$ in (\ref{eq:J_JMhh_def}).
We can regard (\ref{eq:J_JMhh_def}) as a relativistic extension 
of the non-relativistic spinor ${J}_{JM\lambda_n\lambda_p}^{\rm NR}({\bf x};k)$
to the relativistic one ${J}_{JM\lambda_n\lambda_p}({\bf x};k)$,
so that the spinor satisfies the Dirac equation.
We use a notation $J^{\rm NR}|_{\rm R-EX}$ 
for this relativistic extension like as (\ref{eq:J_JMhh_def})
in the follow.

Next we rewrite (\ref{eq:WF_infV_h1}) and (\ref{eq:WF_infV_h2})
by the $(JMls)$-base as
\begin{eqnarray}
&& 
\phi^{\infty}({\bf x};{\bf k})
= 
\sum_{JMls}
C_{JMls}({\bf k}) \cdot 
\phi^{\infty}_{JMls}({\bf x};k)
\ ,
\label{eq:WF_infV_ls1}
\\
&&
\phi^{\infty}_{JMls}({\bf x};k)
= \sum_{l's'} 
\Bigl[
{J}_{JMl's'}({\bf x};k) \cdot \alpha_{l's',ls}^{(J)}(k)  +
{N}_{JMl's'}({\bf x};k) \cdot  \beta_{l's',ls}^{(J)}(k)
\Big]
\ , 
\label{eq:WF_infV_ls2}
\end{eqnarray}
with some constant $C_{JMls}({\bf k})$, 
where functions of the $(JMls)$-base are defined by 
\begin{eqnarray}
&&
{J}_{JMls}({\bf x};k)
= \sum_{\lambda_n\lambda_p}
{J}_{JM\lambda_n\lambda_p}({\bf x};k) 
\cdot \langle JMls | JM\lambda_n\lambda_p \rangle
\ , 
\label{eq:J_JMls_def}
\\
&& 
\alpha_{l's',ls}^{(J)}(k)
=
\sum_{\xi_n \xi_p \lambda_n \lambda_p }
\alpha_{\xi_n\xi_p,\lambda_n\lambda_p }^{(J)}(k) \cdot 
\langle JMl's' | JM     \xi_n     \xi_p \rangle 
\langle JMl s  | JM \lambda_n \lambda_p \rangle
\ , 
\label{eq:alpha_beta_JMls_def}
\end{eqnarray}
and ${N}_{JM\lambda_n\lambda_p}({\bf x};k)$ and 
$\beta_{l's',ls}^{(J)}(k)$ are similarly defined.
Here we should note that
${J}_{JMls}({\bf x};k)$ and 
${N}_{JMls}({\bf x};k)$
are not eigenstates of the orbital angular momentum
and the spin with $l$ and $s$.
These functions satisfy the Dirac equation,
thus the upper and the lower components
have different orbital angular momenta.
%
%
\section{ Wave function on the finite volume }
Next we consider the wave function on the finite periodic box
of volume $L^3$ defined by
\begin{equation}
\phi^{L}_{\alpha\beta}({\bf x};k)
= 
\langle 0 | 
\ n_{\alpha}( {\bf x}/2) 
\ p_{\beta }(-{\bf x}/2) \ | k \rangle
\ , 
\label{eq:WF_finV_def}
\end{equation}
where 
$| k \rangle$ is the energy eigenstate with $\sqrt{s}=2\sqrt{m^2+k^2}$
on the finite volume.
Here we assume the condition $R<L/2$
for the two-nucleon interaction range $R$ and the lattice size $L$,
so that the boundary condition does not distort
the shape of the two-nucleon interaction.
In the region $R<|{\bf x}|<L$,
the wave function satisfies following two equations
and the boundary condition.
\begin{eqnarray}
&&
\left[\ i ({\bf \gamma}\cdot{\bf \nabla} ) 
            + \gamma^0 E - m  \ \right]
\phi^{L}({\bf x};k) 
= 0
\ , 
\quad 
\phi^{L}({\bf x};k)
\left[\ -i ({\bf \gamma}\cdot\stackrel{\leftarrow}{\bf \nabla} ) 
            + \gamma^0 E - m  \ \right]^{\rm T}
= 0
\ , 
\label{eq:WF_finV_DEQ}
\\
&&
\phi^{L}({\bf x}+{\bf n}L ;k) = 
\phi^{L}({\bf x}          ;k) 
\qquad ( \ {\bf n} \in \mathbb{Z}^3 \ ) 
\ ,
\label{eq:WF_finV_BC}
\end{eqnarray}
where $E=\sqrt{m^2+k^2}$.
The general solution of these equations can be written
by the linear combination of the Green function
defined by 
\begin{eqnarray}
&&   
G_{JMls}({\bf x};k) 
= G_{JMls}^{\rm NR}({\bf x};k) \bigl|_{\rm R-EX}
\ ,
\label{eq:G_JMls_def1} 
\\
&&
G_{JMls}^{\rm NR}({\bf x};k)
=
{\cal Y}_{JM}^{ls}(\nabla) \frac{1}{L^3}
\sum_{ {\bf p}\in \Gamma }
\frac{ 1 }{ p^2 - k^2 }
{\rm e}^{ i {\bf p}\cdot{\bf x} } 
\ , \quad
{\cal Y}_{JM}^{ls}({\bf p}) = p^l \cdot Y_{JM}^{ls}(\Omega_p)
\ ,
\label{eq:G_JMls_def2} 
\end{eqnarray}
where
$\Gamma = \{ {\bf p} | 
{\bf p}=(2\pi)/L \cdot {\bf n} \ , \ {\bf n}\in\mathbb{Z}^3 \}$
and $\Omega_p$ is the spherical coordinate for ${\bf p}$.
This Green function is related to 
that introduced in Ref.~\cite{fm_LU} $G_{lm}({\bf x};k)$
by \hfill\break
$G_{JMls}^{\rm NR}({\bf x};k)
= \sum_{m \mu} G_{lm}({\bf x};k)
\cdot \phi(s,\mu) C(lm;s\mu;JM)$.

Using partial wave expansion of $G_{lm}({\bf x};k)$
given in Ref.~\cite{fm_LU},
we obtain
\begin{eqnarray}
G_{JMls}({\bf x};k)
=
a_l(k) b_l(k) \cdot {N}_{JMls}({\bf x};k) 
+ 
a_l(k) 
\sum_{J'M'l'}
b_{l'}(k) \cdot {J}_{J'M'l's}({\bf x};k)
\cdot M_{J'M'l',JMl}^{(s)}(k)
\ , 
\label{eq:G_JMls_pwexp}
\end{eqnarray}
where $a_l(k)=(-1)^l k^{l+1} / (4\pi)$,
$b_l(k)$ is the normalization constant
appeared in (\ref{eq:J_JMls_NR_def_2})
and
\begin{equation}
M_{J'M'l',JMl}^{(s)}(k)
= \sum_{mm' \mu}
M_{l'm',lm}(k) \cdot
C( l' m' ; s \mu ; J' M' )
C( l  m  ; s \mu ; J  M  )
\ .
\label{eq:M_JML_def}
\end{equation}
The function 
$M_{l'm',lm}(k)$ in (\ref{eq:M_JML_def})
is defined by (3.34) in Ref.~\cite{fm_LU},
which is given by
\begin{eqnarray}
&&
M_{l'm',lm}(k)
= \sum_{l''m''} I_{l'm',l''m'',lm} W_{l''m''}(q)
\ , \quad  q= k L / (2\pi)
\\ 
&& 
I_{l'm', l'' m'',lm} 
= 
(-1)^l \cdot i^{l+l'} \cdot (2l''+1) \sqrt{ \frac{ 2l+1 }{ 2l'+1 } } 
\cdot 
C( l 0 ; l'' 0   ; l' 0  )
C( l m ; l'' m'' ; l' m' )
\ , \\
&&
W_{lm}(q) 
= \frac{1}{ \pi^{3/2} q^{l+1} \sqrt{2l+1} }
\ \sum_{ {\bf n} \in \mathbb{Z}^3 } 
\frac{1}{ n^2 - q^2}  {\cal Y}_{lm}({\bf n}) 
\ , \quad 
{\cal Y}_{lm}({\bf n}) = n^l \cdot Y_{lm}(\Omega_n)
\ .
\label{eq:W_lm_def}
\end{eqnarray}
%
%
\section{ Relation between $\phi^{\infty}$ and $\phi^{L}$ }
In the following 
we restrict ourselves to the wave function
for the irreducible representation of the rotational group 
on the finite volume (cubic group ${\rm O}$),
which is defined by 
\begin{equation}
\phi^{L}_{\Gamma\alpha} ({\bf x};k)
=
\langle 0 | 
\ n( {\bf x}/2) 
\ p(-{\bf x}/2)
\ | k ; \Gamma \alpha \rangle
\ , 
\label{eq:WF_finV_IRREP_def}
\end{equation}
where 
$|k ; \Gamma \alpha \rangle$
is the energy eigenstate with $\sqrt{s}=2\sqrt{m^2+k^2}$ and 
belongs to the irreducible representation of ${\rm O}$ 
labeled by $\Gamma$ and $\alpha$
($\alpha=1\dots \dim\Gamma$ , $\Gamma = \{ A_1, A_2, E, T_1, T_2\}$).
Projection of the irreducible representation of SU(2) ($|JM\rangle$)
to that of ${\rm O}$ ($|\Gamma \alpha n J \rangle$)
is given by
\begin{equation}
|JM \rangle
= \sum_{\Gamma \alpha n }
|\Gamma \alpha nJ \rangle \cdot V(JM;\Gamma\alpha nJ)^{*}
\ , \quad
|\Gamma\alpha nJ \rangle
= \sum_{M}
|JM \rangle \cdot V(JM;\Gamma\alpha nJ)
\ , 
\label{eq:SU2_O_Proj}
\end{equation}
with known coefficient $V(JM;\Gamma\alpha nJ)$,
where $n$ is the multiplicity of the representation $\Gamma$.

In the previous section 
the wave functions
were expanded in terms of functions of the $(JMls)$-base
($J_{JMls}$ and $N_{JMls}$).
But it is more convenient 
for the wave function (\ref{eq:WF_finV_IRREP_def})
to expand in terms of
functions of $(\Gamma\alpha nJls)$-base defined by 
\begin{equation}
{J}_{\Gamma\alpha nJls}({\bf x};k)  = 
\sum_{M} J_{JMls}({\bf x};k) \cdot 
V(JM;\Gamma\alpha nJ)
\ ,
\label{eq:Trans_to_IRR_O}
\end{equation}
with the coefficient $V(JM;\Gamma\alpha nJ)$.

In the region $|{\bf x}|>R$,
the wave function (\ref{eq:WF_finV_IRREP_def})
can be written by the linear combination of the Green function
and also the wave function in the infinite volume as
\begin{eqnarray}
\phi^{L}_{\Gamma\alpha} ({\bf x};k)
=
\sum_{nJls}
E_{\Gamma\alpha nJls}({\bf k})
\cdot G_{\Gamma\alpha nJls}({\bf x};k)
= 
\sum_{nJls}
C_{\Gamma\alpha nJls}({\bf k})
\cdot \phi^\infty_{\Gamma\alpha nJls}({\bf x};k)
\ , 
\label{eq:WF_finV_IRREP_exp}
\end{eqnarray}
with some coefficients
$E_{\Gamma\alpha nJls}({\bf k})$ and
$C_{\Gamma\alpha nJls}({\bf k})$, 
where
$G_{\Gamma\alpha nJls}({\bf x};k)$ and 
$\phi^{\infty}_{\Gamma\alpha nJls}({\bf x};k)$
are functions of the $(\Gamma\alpha nJls)$-base obtained
by the transformation (\ref{eq:Trans_to_IRR_O})
from 
$G_{JMls}({\bf x};k)$ defined by (\ref{eq:G_JMls_def1}) and 
$\phi^{\infty}_{JMls}({\bf x};k)$ defind by (\ref{eq:WF_infV_ls2}).
After some calculations we obtain
\begin{eqnarray}
\phi^{L}_{\Gamma\alpha}({\bf x};k)
&=& 
\sum_{nJls}
E_{\Gamma\alpha nJls}({\bf k})
\left(
b_{l}(k) \cdot {N}_{\Gamma\alpha nJls}({\bf x};k) 
+
\sum_{n'J'l'} 
b_{l'}(k) \cdot {J}_{\Gamma\alpha n'J'l's}({\bf x};k) \cdot 
M_{n'J'l',nJl}^{(s)}(\Gamma;k)
\right)
\cr
&=& 
\sum_{nJls} 
C_{\Gamma\alpha nJls}({\bf k})
\sum_{l's'}
\left(
{J}_{\Gamma\alpha nJl's'}({\bf x};k) \cdot \alpha_{l's',ls}^{(J)}(\Gamma;k) +
{N}_{\Gamma\alpha nJl's'}({\bf x};k) \cdot  \beta_{l's',ls}^{(J)}(\Gamma;k)
\right)
\ , \quad 
\label{eq:WF_finV_IRREP_exp_2}
\end{eqnarray}
where the constant $a_l(k)$
are removed by redefinition
of the constant $E_{\Gamma\alpha nJls}({\bf k})$,
and 
\begin{eqnarray}
&&
\delta_{\Gamma'\Gamma}
\delta_{\alpha'\alpha}\cdot
M_{n'J'l',nJl}^{(s)}(\Gamma;k)
=
\sum_{MM'}
M_{J'M'l',JMl}^{(s)}(k) \cdot
V( J'M' ; \Gamma' \alpha' n' J' )
V( J M  ; \Gamma  \alpha  n  J  )
\ , 
\label{eq:M_Gamma_def}
\\
&&
\alpha_{l's',ls}^{(J)}(\Gamma;k) 
=
\sum_{M}
\alpha_{l's',ls}^{(J)}\cdot
V(JM;\Gamma \alpha nJ )
V(JM;\Gamma \alpha nJ )
\ , 
\label{eq:alpha_Gamma_def}
\end{eqnarray}
and $\beta_{l's',ls}^{(J)}(\Gamma;k)$ is similarly defined.
The diagonal property
of $M(\Gamma;k)$ in (\ref{eq:M_Gamma_def})
for indices $(\Gamma\alpha)$ 
is result from
the invariance of $M_{J'M'l',JMl}^{(s)}(k)$
under the rotation on the finite volume
(see Ref.~\cite{fm_LU}).

From (\ref{eq:WF_finV_IRREP_exp_2}),
we know that coefficients of functions
${J}_{\Gamma\alpha nJls}({\bf x};k)$ and 
${N}_{\Gamma\alpha nJls}({\bf x};k)$
relate each other.
After some calculations,
we find that it is given by 
\begin{equation}
\det\left[ 
\ {\bf M}(\Gamma;k) \ - 
\ {\bf A}(\Gamma;k) / {\bf B}(\Gamma;k) \ \right] = 0
\ , 
\label{eq:finitesize_formula}
\end{equation}
where 
we introduce a vector space spanned by indices $(nJls)$
at fixed $(\Gamma\alpha)$ and define 
linear operators on this vector space by 
\begin{equation}
\bigl[{\bf M}(\Gamma;k)\bigr]_{n'J'l's',nJls} =
\delta_{s's} \cdot M_{n'J'l',nJl}^{(s)}(\Gamma;k)
\ , \quad 
\bigl[{\bf A}(\Gamma;k)\bigr]_{nJ'l's',nJls} = 
\delta_{n'n} \delta_{J'J} \cdot 
\alpha_{l's',ls}^{(J)}(\Gamma;k)/b_{l'}(k)
\ , 
\label{eq:M_Gamma_AB_Gamma_defV}
\end{equation}
and ${\bf B}(\Gamma;k)$ is similarly defined.
Equation (\ref{eq:finitesize_formula}) 
is a finite size formula for the elastic $NN$ scattering system,
which gives us
a relation between the energy eigenvalue on the finite volume
and the quantity of the elastic scattering ${\bf A}/{\bf B}$.
%
%
\section{ Finite size formula for $NN$ system }
In this section
we show the explicit matrix form 
of the finite size formula for the $NN$ system
(\ref{eq:finitesize_formula}).
$S$-matrix at fixed $J$ forms a $4\times 4$ matrix.
This matrix is reduced to sub-matrices
by the eigenvalue of 
the global symmetry : 
the parity $P$ and the particle exchange $R$ 
($=(-1)^I$ with the iso-spin $I$) as
\begin{equation}
S^{(J)}
=
\begin{array}[t]
{c  l        l                            c     l              c   l                 l }
  & \Bigl( & \mbox{$2\times 2$\ matrix} & ; \ & P=(-1)^{J-1} & , & R=(-1)^{J-1} & \Bigr) \\
+ & \Bigl( & \mbox{$1\times 1$\ matrix} & ; \ & P=(-1)^{J  } & , & R=(-1)^{J  } & \Bigr) \\
+ & \Bigl( & \mbox{$1\times 1$\ matrix} & ; \ & P=(-1)^{J  } & , & R=(-1)^{J-1} & \Bigr) \\
\end{array}
\ .
\end{equation}
${\bf A}(\Gamma;k)$ and ${\bf B}(\Gamma;k)$ in the finite size formula 
(\ref{eq:finitesize_formula})
also take same form.

We note that 
the basis of the partial wave expansion
${J}_{\Gamma\alpha nJls}({\bf x};k)$ and 
${N}_{\Gamma\alpha nJls}({\bf x};k)$
in (\ref{eq:WF_finV_IRREP_exp_2})
are eigenstates of
the parity and the particle exchange
with $P=(-1)^l$ and $R=(-1)^l \cdot (-1)^{s-1}$.
Thus 
the wave function for the state with $R=-P$,
only functions with $s=0$ appear
in the partial wave expansion.
For the state with $R=P$,
only functions with $s=1$ appear.
The mixing between $s=0$ and $s=1$ is forbidden
by the symmetry of the parity and the particle exchange (iso-spin).
Therefore we can separately
obtain the finite size formula
for $P=-R$ ($s=0$) and $P=R$ ($s=1$).

In the case of $R=-P$ ($s=0$),
the components of the matrix 
${\bf M}(\Gamma;k)$
in the finite size formula 
(\ref{eq:finitesize_formula})
are given by
\begin{equation}
M_{n'J'l',nJl}^{(s)}(\Gamma;k)
=
\delta_{J'l'} \delta_{Jl} \cdot 
\sum_{MM'}
M_{J'M',JM}(k) \cdot
V( \Gamma\alpha nJ' ; J'M' )
V( \Gamma\alpha nJ  ; J M  )
\ .
\label{eq:M_s0}
\end{equation}
This is the same matrix 
as that appeared in the finite size formula
for the two-meson system.
Further,
$\alpha^{(J)}_{l's,ls}(k)
= \delta_{Jl}\delta_{Jl'} \cdot \alpha_l(k)$ and
$\beta^{(J)}_{l's,ls}(k)
= \delta_{Jl}\delta_{Jl'} \cdot \beta_l(k)$
with the diagonal components $\alpha_l(k)$ and $\beta_l(k)$
also take same matrix form as that for the two-meson system.
Thus the finite size formula 
for the $NN$ system with $P=-R$ ($s=0$) is same as 
that for the two-meson system in Ref.~\cite{fm_LU}.

In the case of $R=P$ ($s=1$),
the matrix
${\bf M}(\Gamma;k)$ and ${\bf A}(\Gamma;k)/{\bf B}(\Gamma;k)$
have complicated structure.
In the following
we show explicit matrix form of the finite size formula 
for some channels as example.
We neglect contributions of $J\ge 5$.
In this case the multiplicity $n$ 
is 1 for all irreducible representations $\Gamma$, 
thus we omit the index $n$ in the formula
(\ref{eq:finitesize_formula}) for simplicity,
as
\begin{eqnarray}
&&
\det\left[
\ {\bf M}(\Gamma;k) \ - 
\ {\bf A}(\Gamma;k)/{\bf B}(\Gamma;k) \ \right] = 0 
\ , 
\label{eq:finitesize_formula_s1}
\\
&&
\bigl[{\bf M}(\Gamma;k)\bigr]_{J'l',Jl} =
M_{n'J'l',nJl}^{(s)}(\Gamma;k)
\ , \quad 
\bigl[{\bf A}(\Gamma;k)\bigr]_{J'l',Jl} = 
\delta_{J'J} \cdot \alpha_{l's',ls}^{(J)}(\Gamma;k)/b_{l'}(k) 
\ , 
\label{eq:M_Gamma_AB_Gamma_defV_s1}
\end{eqnarray}
where $n=n'=1$, $s=s'=1$ and
the matrix ${\bf B}(\Gamma;k)$ is similarly defined.

The first example is the deuteron state.
We have to consider the $NN$ state with the total angular momentum $J=1$
and the parity $P=+1$,
which corresponds to
${}^3S_1$ and ${}^3D_1$ states
in the non-relativistic limit.
The $J=1$ state belongs to the irreducible representation 
of the cubic group $\Gamma=T_1$,
thus we consider the finite size formula 
for $\Gamma=T_1$ for the study of the deuteron.
The other angular momentum states also belong to $T_1$
as $T_1 = 1 + 3 + 4$ up to $J \ge 5$ and 
the finite size formula includes contributions from all these states.
For each values of $J$, possible values of $l$ are given by
\begin{equation}
\begin{array}{lc}
l = 0 , \ 2  & \ \mbox{ for \ $J=1$ } \\ 
l = 2 , \ 4  & \ \mbox{ for \ $J=3$ } \\
l = 4        & \ \mbox{ for \ $J=4$ }
\label{eq:l_deuteron}
\end{array}
\ , 
\end{equation}
from the parity conservation
and the theory of addition of the angular momentum.
Thus matrices
${\bf M}(\Gamma;k)$ and
${\bf A}(\Gamma;k)/{\bf B}(\Gamma;k)$
in the finite size formula (\ref{eq:finitesize_formula_s1})
take : 
\begin{eqnarray}
&& {\bf M}=
\left(
\begin{array}{lllll}
M_{10,10} & M_{10,12} & M_{10,32} & M_{10,34} & M_{10,44} \\ 
M_{12,10} & M_{12,12} & M_{12,32} & M_{12,34} & M_{12,44} \\ 
M_{32,10} & M_{32,12} & M_{32,32} & M_{32,34} & M_{32,44} \\ 
M_{34,10} & M_{34,12} & M_{34,32} & M_{34,34} & M_{34,44} \\ 
M_{44,10} & M_{44,12} & M_{44,32} & M_{44,34} & M_{44,44} \\ 
\end{array}
\right)
, \quad
{\bf A}/{\bf B}=
\left( \ \begin{array}{ccccc}
\cline{1-2}
\multicolumn{2}{|c|}{\multirow{2}{*}{\ \ \it J=1 \ }} & \ \ 0 & 0 & 0 \\ 
\multicolumn{2}{|c|}{}                                & \ \ 0 & 0 & 0 \\ 
\cline{1-2}
\cline{3-4}
0 & 0 & \multicolumn{2}{|c|}{\multirow{2}{*}{\ \ \it J=3 \ }} & 0 \\
0 & 0 & \multicolumn{2}{|c|}{}                                & 0 \\
\cline{3-4}
\cline{5-5}
0 & 0 & 0 & 0 & \multicolumn{1}{|c|}{\multirow{1}{*}{\it J=4}} \\
\cline{5-5}
\end{array} \ \right)
\ , 
\label{eq:eq:finitesize_formula_D}
\end{eqnarray}
where 
the boxes in the matrix ${\bf A}/{\bf B}$ which enclose the values of $J$
refer to $2\times 2$ or $1\times 1$ matrices
expanded by the possible values of $l$.
In (\ref{eq:eq:finitesize_formula_D})
components of the matrix ${\bf M}$
are denoted by $M_{J'l',Jl} \equiv [{\bf M}(\Gamma;k)]_{J'l',Jl}$
and are given by 
\begin{eqnarray}
&& \quad
\begin{array}[b]{l}
M_{10,10}=W_{00}    \\
M_{12,10}=0         \\
M_{32,10}=0         \\
M_{34,10}=-2 W_{40} \\
M_{44,10}=\frac{6}{7}\sqrt{7} W_{40}  \\
\end{array}
\quad
\begin{array}[b]{l}
M_{12,12}=W_{00}       \\
M_{32,12}=-\frac{6}{7}\sqrt{6 }W_{40}  \\
M_{34,12}=-\frac{5}{7}\sqrt{2 }W_{40}  \\
M_{44,12}=-\frac{3}{7}\sqrt{14}W_{40}  \\
\end{array}
\quad
\begin{array}[b]{l}
M_{32,32}=W_{00} + \frac{6}{7}W_{40}  \\
M_{34,32}=\frac{30}{77}\sqrt{3 }W_{40} + \frac{50}{33}\sqrt{3 }W_{60} \\
M_{44,32}=\frac{18}{77}\sqrt{21}W_{40} + \frac{10}{11}\sqrt{21}W_{60} \\
\end{array}
\cr\cr
&&
\quad
\begin{array}[b]{l}
M_{34,34}=W_{00} + \frac{81}{77}W_{40} + \frac{25}{33}W_{60}  \\
M_{44,34}=-\frac{27}{77}\sqrt{7}W_{40} - \frac{15}{11}\sqrt{7}W_{60} \\
\end{array}
\quad
\begin{array}[b]{l}
M_{44,44}=
                  W_{00} 
+ \frac{81  }{143}W_{40} 
+ \frac{1   }{55 }W_{60} 
+ \frac{1792}{715}W_{80}
\end{array}
\ , 
\end{eqnarray}
where 
$M_{J'l',Jl}= M_{Jl,J'l'}$
and the function $W_{lm}$ is defined by (\ref{eq:W_lm_def}).

Finally we consider the state with same $J$ 
but opposite parity to the deuteron,
{\it ie.} $J=1$, $P=-1$.
We also consider the representation $\Gamma=T_1$.
Possible values of $l$ for each $J$ are different 
from those of the deuteron case as
\begin{equation}
\begin{array}{lc}
l = 1       & \ \mbox{ for $J=1$ } \\ 
l = 3       & \ \mbox{ for $J=3$ } \\
l = 3, \ 5  & \ \mbox{ for $J=4$ }
\label{eq:l_Pdeuteron}
\end{array}
\ .
\end{equation}
Thus matrices in the finite size formula 
take different forms : 
\begin{eqnarray}
&& {\bf M} =
\left(
\begin{array}{lllll}
M_{11,11} & M_{11,33} & M_{11,43} & M_{11,45} \\
M_{33,11} & M_{33,33} & M_{33,43} & M_{33,45} \\
M_{43,11} & M_{43,33} & M_{43,43} & M_{43,45} \\
M_{45,11} & M_{45,33} & M_{45,43} & M_{45,45} \\
\end{array}
\right)
, \quad
{\bf A}/{\bf B}  =
\left( \
\begin{array}{ccccc}
\cline{1-1}
\multicolumn{1}{|c|}{\multirow{1}{*}{\it J=1}} & 0 & 0 & 0 \\
\cline{1-1}
\cline{2-2}
0  & \multicolumn{1}{|c|}{\multirow{1}{*}{\it J=3}} & 0 & 0 \\
\cline{2-2}
\cline{3-4}
0 & 0 & \multicolumn{2}{|c|}{\multirow{2}{*}{\ \ \it J=4 \ }} \\
0 & 0 & \multicolumn{2}{|c|}{}                                \\
\cline{3-4}
\end{array}\ \right)
\ ,
\\
\cr
&&
\quad
\begin{array}[b]{l}
M_{11,11}=W_{00}   \\
M_{33,11}= \frac{3}{7}\sqrt{14 }W_{40} \\
M_{43,11}=-\frac{1}{7}\sqrt{210}W_{40} \\
M_{45,11}= \frac{2}{7}\sqrt{42 }W_{40} \\
\end{array}
\quad
\begin{array}[b]{l}
M_{33,33}=W_{00}  + \frac{3}{11}W_{40} - \frac{25}{11}         W_{60} \\
M_{43,33}=-\frac{3}{11}\sqrt{15}W_{40} - \frac{35}{33}\sqrt{15}W_{60} \\
M_{45,33}= \frac{6}{11}\sqrt{ 3}W_{40} + \frac{70}{33}\sqrt{ 3}W_{60} \\
\end{array}
\cr
&&
\quad
\begin{array}[b]{l}
M_{43,43}=W_{00} + \frac{9}{11}W_{40} - \frac{5}{33}W_{60} \\
M_{45,43}=
  \frac{18 }{143}\sqrt{5}W_{40} 
- \frac{14 }{165}\sqrt{5}W_{60} 
- \frac{896}{715}\sqrt{5}W_{80} \\
\end{array}
\cr
&&
\quad
\begin{array}[b]{l}
M_{45,45}=W_{00} 
+ \frac{126}{143}W_{40} 
- \frac{32 }{165}W_{60} 
- \frac{448}{715}W_{80}
\end{array}
\ . 
\end{eqnarray}
%
%
\section{ Summary }
The finite size formula
for the elastic $NN$ scattering system is derived
from the relativistic quantum field theory.
The extension to other two-baryon system as the $N\Lambda$ system
is trivial.
Finally I give a comment for the $N\pi$ system.
The formulation of this paper for the $NN$ system 
is also valid for the system 
with the general value of the spin $s$.
Thus the finite size formula for the $N\pi$ system 
can be easily obtained from that for the $NN$ system
(\ref{eq:finitesize_formula})
by set $s=1/2$.
In calculations of matrices
${\bf M}(\Gamma;k)$,
${\bf A}(\Gamma;k)$ and 
${\bf B}(\Gamma;k)$ in (\ref{eq:M_Gamma_AB_Gamma_defV}),
we change
the coefficient $V(JM;\Gamma\alpha nJ)$ in (\ref{eq:SU2_O_Proj}) 
by that for the double covered cubic group (${}^2{\rm O}$)
to deal with the half integer value 
of the total angular momentum $J$
as discussed in Ref.~\cite{fm_BLMR}.
I confirmed that
my results are consistent with
those obtained from the non-relativistic effective theory
by Bernard {\it et al.}~\cite{fm_BLMR}.

This work is supported in part by 
Grants-in-Aid of the Ministry of Education
(No.20540248).
%
%

%
%
\end{document}